\documentclass[aps,prl,twocolumn,showpacs,superscriptaddress]{revtex4}

\usepackage{graphicx}
\usepackage{dcolumn}
\usepackage{bm}
\usepackage{amssymb}
\begin{document}


\title{Galaxy rotation curves in de Sitter space}

\author{Maurice H.P.M. van Putten}
\affiliation{Astronomy and Space Science, Sejong University, 98 Gunja-Dong Gwangin-gu, Seoul 143-747, Korea\\}

\date{\today}

\begin{abstract}
Dark energy inferred from the observed negative deceleration parameter introduces a small mass of 
the graviton, that satisfies the Higuchi stability condition. It implies an infra-red modification of gravitation 
that produces Milgrom's inverse distance law of gravitational attraction in excellent agreement with the 
observed galaxy rotation curves. We conclude that dark matter is present cosmologically with no need
for local clustering in galaxies. 
\end{abstract}

\maketitle

\section{Introduction}

General relativity gives an accurate description of gravitational attraction in systems on the scale of our 
solar system (e.g. \cite{bin87}), in a four-covariant embedding of Newton's theory of gravitation by a 
mixed elliptic-hyperbolic system of equations (e.g. \cite{van96}), parameterized by Newton's constant 
$G$ and the velocity of light $c$. 

Modern cosmology, however, points to a mysterious cosmological constant $\Lambda$
or dark energy and dark matter with, at present, no microphysical origin \citep{rie98,per99,rie04}. 
They appear at weak gravity on the cosmological scale of acceleration $a_H=cH_0$, defined by
the Hubble constant $H_0$ and the velocity of light $c$. Assuming Newton's law of gravitation,
dark matter appears in galaxy clusters (by a factor of about eight \citep{gio09}), in the Faber \& Jackson
\citep{fab76} and Tully-Fisher \cite{tul77} relations for galaxy stellar velocity dispersion and, respectively, 
 rotation velocities, in globular clusters \citep{her13} and in ultra-wide stellar binaries \citep{her12}.

The cosmological scale $a_H$ represents the surface gravity of the cosmological event horizon that,
in a three-flat universe, has a Hubble radius $R_H=c/H_0$. Weak gravity, therefore, is conceivably
affected by thermodynamic properties of the cosmological event horizon, which would ordinarily
be negligible when considering gravitation in the solar system. It has a de Sitter temperature \citep{gh77} 
\begin{eqnarray}
k_BT_{dS}=\frac{H\hbar}{2\pi},
\label{EQN_RH}
\end{eqnarray}
where $k_B$ denotes the Boltzmann constant. As a null-surface, it further has a Bekenstein-Hawking entropy 
$S/k_B = (1/4)A_H/l_p^2$. In a holographic interpretation \citep{tho93,sus95}, this is identified with the number of degrees of freedom in the 
phase space of the visible universe which, at present, corresponds to the information required for encoding the microphysical distribution of matter \citep{van14}. By (\ref{EQN_RH}), the cosmological event horizon can hereby be attributed a two-dimensional pressure which, by its Gauss curvature, $R_H^{-2}$, introduces a finite negative pressure, $p_0$, in the enclosed space-time of the visible universe
along with dark energy $\rho_\Lambda=-p_0$ by Lorentz invariance (e.g. \cite{wei89}). By scaling, we infer from
the unit of gravitational luminosity 
\begin{eqnarray}
L_0=\frac{c^5}{G}
\label{EQN_L0}
\end{eqnarray}
a dark energy density
\begin{eqnarray}
\rho_{\Lambda} =  \frac{L_0}{cA_H}: ~~\Omega_\Lambda=\frac{2}{3},
\label{EQN_OML}
\end{eqnarray}
where $\Omega_\Lambda=\rho_\Lambda/\rho_c$, $\rho_c = 3c^2H_0^2/8\pi G$.
The same result can be seen to derive from entropy forces \citep{eas11} by virtual displacements of null-surfaces
following Gibbs' procedure (e.g. \cite{van12}) 
\begin{eqnarray}
\rho_\Lambda = -p = A_H^{-1} T_{dS} \frac{dS}{dR} = \frac{k_BT_{dS}}{2R_Hl_p^2},
\label{EQN_rL}
\end{eqnarray}  
where $l_p=\sqrt{G\hbar/c^3}$ denotes the Planck length.

Here, we point out that a positive dark energy (\ref{EQN_OML}) - as currently observed or inferred from a holographic principle - implies 
a finite mass $m_0=\sqrt{\Lambda}\hbar/c$ to the graviton, as may be seen by coupling to the Ricci tensor in the nonlinear wave equations for the Riemann-Cartan connections \citep{van96}. It hereby obtains a dispersion relation 
\begin{eqnarray}
\omega = c \sqrt{k^2+\Lambda},
\label{EQN_1}
\end{eqnarray}
of energy $\hbar\omega$ in terms of momentum $\hbar k$ at a wave number $k$. This relation is preserved
in when preserving gauge invariance by including Stueckelberg fields (\cite{rue04,aka14} and
references therein). It implies a rest mass energy $\epsilon_0 = \hbar c \sqrt{\Lambda}$ of the graviton. 
The problem of consistent general relativity with massive gravitons has recently received considerable attention (e.g. \cite{der11,ber14}). 
With $-q_0H^2=H^2+\dot{H}$, the generalized Higuchi constraint $m^2 \ge 2 (H^2 + \dot{H})$ \citep{hig87,des01,gri10} reduces to $\Omega_\Lambda \ge -\frac{2}{3}q_0.$ Based on observations, $-1 < q_0< -0.5$ \citep{rie04,wu08,gio12}, whereby $q_0>-1$ appears secure. 

\section{Normalized vacuum temperature}

In light of (\ref{EQN_RH}), the gravitons are warm and assume a non-relativistic temperature. With $k_BT_0=m_0c^2$, we have $\beta_{dS}=T_0/T_{dS}$ satisfying
\begin{eqnarray}
\beta_{dS}= {2\pi\sqrt{2}}.
\label{EQN_3a}
\end{eqnarray}
To an inertial observer, $T_{dS}$ appears in the form of isotropic radiation coming from all directions. 
An accelerating observer experiences additional radiation at an Unruh temperature $k_BT_U=a\hbar/(2\pi c)$ 
of with momenta along the direction of acceleration by equivalence in Rindler and Schwarzschild space times \citep{unr76,hig98}. The momenta of Unruh radiation, $p_U=k_BT_U/c$ and the isotropic momenta $p_{dS}=k_BT_{dS}/c$ of de Sitter background radiation are hereby uncorrelated. The average net momentum in magnitude hereby satisfies 
$p=\sqrt{p_U^2+p_{dS}^2}$, giving rise to an apparent net temperature \citep{nar96,des97,jac98}
\begin{eqnarray}
\hat{T}=\sqrt{1+\hat{T}_U^2},
\label{EQN_3b}
\end{eqnarray}
where the hat refers to normalization with respect to $T_{dS}$. The associated radiation energy $\epsilon=k_BT$ hereby satisfies the same relation as (\ref{EQN_1}) with $k_BT_{dS}$ setting a minimum temperature \citep{kli11}. 

In holography applied to two-dimensional screens surrounding a region of space of interest, the screen 
temperature would normally satisfy
\begin{eqnarray}
T=\left( \frac{\partial S}{\partial E}\right)^{-1},
\label{EQN_4a}
\end{eqnarray}
where $E$ denotes the enclosed total energy and $S$ is the entropy of the screen defined by 
its phase space. Neglecting $T_{dS}$, it recovers $T_U$ above with Newton's law of 
gravitation as an entropic force \citep{ver11}. The latter can be obtained from Gibbs' principle 
applied to deformations of light cones and the apparent event horizons of black holes alike
endowed with the Bekenstein-Hawking entropy \citep{van12}.

We next apply Gibbs' principle to holography in a de Sitter background.
In the imaging of a particle, let 
\begin{eqnarray}
E=m_0c^2 + e
\label{EQN_e}
\end{eqnarray}
denote the enclosed mass-energy in terms of an associated rest mass energy $m_0c^2$ and internal energy $e$
attributed to the encoding in the screen. By $e$, the screen temperature raises above $T_{dS}$ according to
\begin{eqnarray}
[T]=\left( \frac{\partial S}{\partial e}\right)^{-1},
\label{EQN_4b}
\end{eqnarray}
where $[T]=T-T_{dS}$. We consider holographic imaging by excitation of on-shell gravitational modes 
satisfying (\ref{EQN_1}) at the temperature (\ref{EQN_4b}). Consequently, $e=mc^2/(1+\beta)$ and 
$m_0c^2 =\beta m c^2/(1+\beta)$, where $\beta=T_0/T$ generalizes (\ref{EQN_3a}). For a spherical screen of finite radius $r$ with $N=4\pi r^2/l_p^2$ Planck sized surface elements, (\ref{EQN_4b}) implies
\begin{eqnarray}
e=\frac{1}{2}N[k_BT].
\label{EQN_5}
\end{eqnarray}
In the limit of arbitrarily large $r$, $[k_BT]$ approaches zero and $\beta$ increases to its maximum $\beta_{dS}$ in (\ref{EQN_3b}).

The holographic formulation (\ref{EQN_3b}) and (\ref{EQN_5}) implicitly defines $T_U$ as a function of $N$. Here, 
$N$ is expressed in terms of $T_N=a_N\hbar/(2\pi c)$, parametrized by the 
Newtonian acceleration $a_N=Gm/r^2$ around an enclosed baryonic mass $m$ in the approximation of spherical
symmetry. Corrections for realistic galaxies depend on the ratio of mass in the central region and the disk. For minor
corrections at large distances to the latter, see \cite{mcg98}. Explicitly, 
\begin{eqnarray}
\hat{T}_U=\frac{1}{2}\tilde{T} + \frac{1}{2}\sqrt{\tilde{T}^2+4\beta_{dS}},~~\tilde{T} = \hat{T}_N+1-\beta_{dS}.
\label{EQN_6}
\end{eqnarray} 

\section{Gravitational acceleration}

Fig. 1 shows the observed acceleration $a=2\pi k_BT_U c/\hbar$ as a function of $a_N$ 
and current data on galaxy rotation curves, where the abscissa is the normalized acceleration $a/a_{H}$, $a_H=H_0c$.
Fig. 1 shows a high $\beta$ limit at low-acceleration, known as Milgrom's law \citep{mil83,fam12},
\begin{eqnarray}
a_+= \sqrt{a_Na_0}~~(a_N<<a_H),
\label{EQN_7}
\end{eqnarray}
where
\begin{eqnarray}
a_0 = a_H\frac{2}{1+\beta_{dS}}\simeq 1.37\times 10^{-8}\,\mbox{cm s}^{-2}.
\label{EQN_8}
\end{eqnarray}
The low $\beta$ limit for which $e \simeq (1/2)Nk_BT$ gives
\begin{eqnarray}
a_-=a_N  - 2\pi\sqrt{2}\,a_H+O(a_H^2/a),
\label{EQN_8b}
\end{eqnarray}
where the leading order term is the Newtonian gravitational attraction of \citep{ver11}.
Our model (\ref{EQN_6}) shows a minimum $a/a_N\simeq 0.32$ at $a_N/a_H\simeq 4.6$, when $T_U$ is
similar to $T_{dS}$. The corresponding radius is about
\begin{eqnarray}
r_b \simeq  \frac{1}{2} \sqrt{R_b R_H} = 2.17\,  M_{11}^\frac{1}{2} \, \mbox{kpc}
\label{EQN_9}
\end{eqnarray}
for a galaxy of mass $M=M_{11} 10^{11} M_\odot$, where $a_N\simeq 3\times 10^{-7}$ cm s$^{-2}$. 
Milgrom and Newton's law hold asymptotically in, respectively, $r>>r_0$ and $r<<r_0$. Around
a stellar mass object of mass $M$, (\ref{EQN_9}) may alternatively be expressed as
$r_b \simeq  1.41\times 10^3\,  (M/M_\odot)^\frac{1}{2}$ AU.

\begin{figure}
\centerline{\includegraphics[width=90mm,height=80mm]{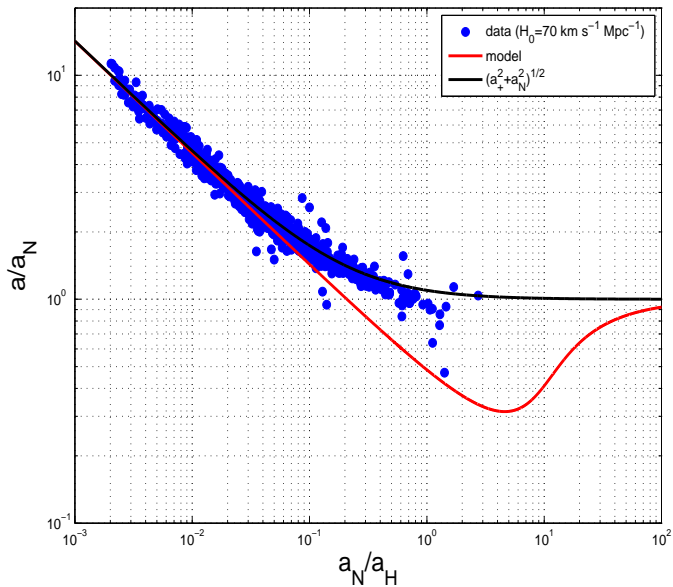}}
\caption{The observed acceleration $a/a_N$ in galaxy rotating curves as a function of the 
Newtonian acceleration $a_N/a_H$ based on baryonic matter ({\em dots}).
(Data courtesy of  \citep{fam12}.) For $a<<a_0$, the model ({\em red curve, $q_0=-1$}) 
describes $a\propto a_N^{q}$ with $\alpha=1/2$ close to $\alpha=0.49$ in a best-fit asymptote 
to the data ({\em black line}) within about bout 10\%. The limit $a>>a_0$ recovers Newton's law. 
To guide the eye, a graph of $\sqrt{a^2+a_H^2}$ is included. The model describes a transition region
around $r_0$ (\ref{EQN_9}) of relatively weak gravitational attraction, where the approximation
of spherical symmetry may be breaking down.}
\label{fig:2}
\end{figure}

\section{Conclusions}
By the observed presence of dark energy, we have a non-trivial dispersion relation of the graviton at extremely low energies.
It implies an enhancement in coupling to matter in the domain where gravitation is weak (high $\beta$), relative to the Hubble scale $a_H$. 
While the low $\beta$ limit recovers Newton's law (\ref{EQN_8b}) when gravitation is within (\ref{EQN_9}), the high $\beta$ limit
recovers the asymptotic behavior in the observed galaxy rotation curves in accord with Milgrom's law, here with a specific
model prediction (\ref{EQN_8}) with no free parameters or fine-tuning.

Our model for galaxy rotation curves shows no need for local concentrations of dark matter. A cosmological distribution of dark
matter may be present that, like dark energy, is possibly uniformly distributed associated with the cosmological event horizon.
Based on \cite{cai05}, it can be shown that the de Sitter temperature $T_{dS}$ vanishes in an early radiation dominated,
when the deceleration parameter equals one. Consequently, the present arguments leave nucleosynthesis unaffected. 

{\bf Acknowledgments.} The author gratefully thanks S.S. McGaugh for the data in Fig. 1.

\end{document}